\documentclass[conference]{IEEEtran}
\IEEEoverridecommandlockouts
\usepackage{cite}
\usepackage{amsmath,amssymb,amsfonts}
\usepackage{algorithmic}
\usepackage{graphicx}
\usepackage{textcomp}
\usepackage{xcolor}
\usepackage{algorithm}
\usepackage{amsfonts}
\usepackage{cite}
\usepackage{graphicx}
\usepackage{amssymb}
\usepackage{caption}
\usepackage{epsfig}
\usepackage{subfigure}
\usepackage{amsmath}
\usepackage{epstopdf}
\newtheorem{theorem}{Theorem}

\usepackage{array}
\usepackage{color}
\usepackage{amsmath}
\usepackage{longtable}
\usepackage{algorithmic}
\usepackage{algorithm}
\usepackage{cases}
\usepackage{mathrsfs}
\usepackage{psfrag}
\usepackage{url}
\usepackage{setspace}

\usepackage{etoolbox}
\makeatletter
\patchcmd{\@makecaption}
  {\scshape}
  {}
  {}
  {}
\makeatletter
\patchcmd{\@makecaption}
  {\\}
  {.\ }
  {}
  {}
\makeatother

\newtheorem{Prob}{Problem}

\newtheorem{remark}{Remark}

\newcounter{TempEqCnt}

\IEEEoverridecommandlockouts

\newcommand{\blue}[1]{\textcolor{black}{#1}}

\def\BibTeX{{\rm B\kern-.05em{\sc i\kern-.025em b}\kern-.08em
    T\kern-.1667em\lower.7ex\hbox{E}\kern-.125emX}}
\begin{document}

\title{Low-complexity Robust Optimization for an IRS-assisted Multi-Cell Network\\
}

\author{\IEEEauthorblockN{Yuhang Jia}
\IEEEauthorblockA{Shanghai Jiao Tong Univ., China}
\IEEEauthorblockA{Jay\_Yoga@sjtu.edu.cn}
\and
\IEEEauthorblockN{Wuyang Jiang}
\IEEEauthorblockA{Shanghai Univ. of Engineering Science, China}
\IEEEauthorblockA{jiang-wuyang@sues.edu.cn}
\and
\IEEEauthorblockN{Ying Cui}
\IEEEauthorblockA{Shanghai Jiao Tong Univ., China}
\IEEEauthorblockA{cuiying@sjtu.edu.cn}
\thanks{This work was supported in part by the National Key R$\&$D Program of China under Grant 2018YFB1801102 and Natural Science Foundation of Shanghai under Grant 20ZR1425300. This work is going to appear in 2021 IEEE GLOBECOM.
}
}

\maketitle

\begin{abstract}
The impacts of channel estimation errors, inter-cell interference, phase adjustment cost, and computation cost on an intelligent reflecting surface (IRS)-assisted system are severe in practice but have been ignored for simplicity in most existing works. In this paper, we \blue{investigate} a multi-antenna base station (BS) serving a single-antenna user with the help of a multi-element IRS in the presence of channel estimation errors and inter-cell interference. \blue{We consider imperfect channel state information (CSI) at the BS, i.e., imperfect CSIT, and focus on the robust optimization of the BS’s instantaneous CSI-adaptive beamforming and the
IRS’s quasi-static phase shifts.} First, we formulate the robust optimization of the BS's instantaneous channel state information (CSI)-adaptive beamforming and IRS's quasi-static phase shifts for the ergodic rate maximization as a very challenging two-timescale stochastic non-convex problem. Then, we obtain a closed-form beamformer for any given phase shifts and a more tractable single-timescale stochastic non-convex problem only for phase shifts. Next, we propose a low-complexity stochastic algorithm to obtain quasi-static phase shifts which correspond to a KKT point of the single-timescale stochastic problem. It is worth noting that the proposed method offers a closed-form robust \blue{instantaneous CSI-adaptive} beamforming design that can promptly adapt to \blue{rapid} CSI \blue{changes over slots} and a robust quasi-static phase shift design of low computation and phase adjustment costs in the presence of channel estimation errors and inter-cell interference. Finally, numerical results demonstrate the notable gains of the proposed robust joint design over existing \blue{ones and reveal the practical values of the proposed solutions}.
\end{abstract}


\section{Introduction}
Recently, intelligent reflecting surface (IRS), consisting of nearly passive, low-cost, reflecting elements with reconfigurable parameters, has been   envisioned to serve as a promising solution for improving spectrum and energy efficiency and has received more and more attention.
Most existing works consider an IRS-assisted system with one multi-antenna base station (BS) serving one or multiple users with the help of one multi-element IRS and optimize BS beamforming and phase shifts of the IRS \cite{QingqingWu2,SJin2,YuhangJia,CGuo}.
In \cite{QingqingWu2}, both the beamformer of
the BS and the phase shifts of the IRS adapt to instantaneous channel state information (CSI).
In \cite{SJin2,YuhangJia,CGuo}, the BS beamformer adapts to instantaneous CSI as in \cite{QingqingWu2}, while the phase shifts of the IRS are adaptive to the statistics of CSI, which remains unchanged over several slots \cite{SJin2,YuhangJia,CGuo}, in contrast with \cite{QingqingWu2}.
Compared with an instantaneous CSI-adaptive phase shift design, a quasi-static phase shift design yields a low phase adjustment cost at the sacrifice of some performance. Considering the practical implementation issue, a quasi-static phase shift design may be more valuable.


Note that most BS beamforming designs \cite{QingqingWu2,SJin2,YuhangJia,CGuo} rely on perfect instantaneous CSI of the cascaded channel and direct channel at the BS, i.e.,
perfect instantaneous channel state information at the transmitter (CSIT).  As channel estimation
in an IRS-assisted system is more challenging than in a conventional system without IRS,
estimation errors are inevitable in practice. The authors in \cite{xu2020resource,hong2020robust} investigate the robust optimization of beamforming and phase shifts to maximize the worst-case average sum rate \cite{xu2020resource} and minimize the transmit power \cite{hong2020robust}, under imperfect instantaneous CSIT. The resulting non-convex robust optimization problems are generally more challenging than their counterparts under perfect instantaneous CSIT. Rather than directly tackling the challenging robust optimization problems, the authors in \cite{xu2020resource,hong2020robust} consider
their simplified versions and propose iterative algorithms to obtain locally optimal solutions
or KKT points. The convergence speeds of the iterative algorithms for
imperfect instantaneous CSIT in \cite{xu2020resource,hong2020robust} are lower than those for perfect instantaneous CSIT in \cite{QingqingWu2,SJin2,YuhangJia,CGuo}.    Furthermore, the robust phase shift designs in \cite{xu2020resource,hong2020robust} rely on the imperfect instantaneous CSIT and hence have higher
phase adjustment costs and are less practical. Therefore, it is critical to obtain robust instantaneous CSI-adaptive beamforming designs with highly efficient methods and robust quasi-static phase shift designs with low phase adjustment costs.

Furthermore, notice that most existing works, including the abovementioned ones under perfect CSIT \cite{QingqingWu2,SJin2,YuhangJia,CGuo} and imperfect CSIT \cite{xu2020resource,hong2020robust}, consider single-cell networks and ignore interference from other BSs. However, in practice, inter-cell interference usually has a severe impact, especially for dense networks or cell-edge users. It is thus critical to take into account the influence of interference when designing practical IRS-assisted systems. In \cite{CPan1,QingqingWu3,YuhangJia}, the authors propose instantaneous CSI-adaptive beamforming designs and instantaneous CSI-adaptive phase shift designs \cite{CPan1,QingqingWu3} or quasi-static phase shift designs \cite{YuhangJia} for IRS-assisted multi-cell networks with inter-cell interference. As \cite{CPan1,QingqingWu3,YuhangJia} assume perfect instantaneous CSIT, the proposed solutions for multi-cell networks are not robust against channel estimation errors. Thus, it is highly desirable to obtain robust beamforming and phase shift design for IRS-assisted multi-cell networks with inter-cell interference.

In this paper, we shall address the above issues. Specifically, we consider a multi-antenna BS serving a single-antenna user with the help of a multi-element IRS in a multi-cell network with inter-cell interference.  The indirect signal and interference links passing the IRS
are modeled with Rician fading, whereas the direct signal and interference links follow Rayleigh fading. We also consider channel estimation errors, phase adjustment cost, and computation cost. Therefore, we focus on the robust optimization of the BS's instantaneous CSI-adaptive beamforming and IRS's quasi-static phase shifts. First, we formulate a robust optimization problem to maximize the user's ergodic rate. Such problem is a very challenging two-timescale stochastic non-convex problem, as the beamforming design and phase shift design are in different time-scales and the number of random variables involved is prohibitively high. Then, by analyzing the expectations of the received signal power and interference
power and by exploiting structural properties, we obtain a closed-form beamformer for any given phase shifts and a more tractable single-timescale stochastic non-convex problem only for phase shifts. Next, using stochastic
successive convex approximation (SSCA), we propose a low-complexity
iterative algorithm, which relies on a closed-form solution to an approximate problem in each iteration, to obtain the quasi-static  phase shifts corresponding to a KKT point of
the single-timescale stochastic non-convex problem. Notice that the proposed method is highly desirable in practice. Because it does not require computing or adjusting phase shifts in each slot, and it is effective under channel estimation errors and inter-cell
interference. Finally, numerical results demonstrate notable
gains of the proposed method over existing instantaneous CSI-adaptive beamforming and quasi-static phase shift designs.

\section{System Model}\label{sec:system}
\begin{figure}[t]
\begin{center}
 \includegraphics[width=4.5cm]{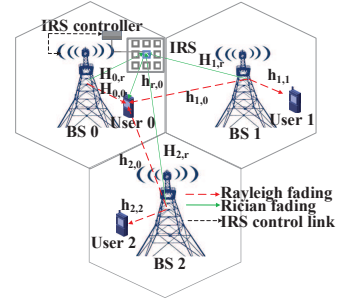}
  \end{center}
 \vspace*{-2mm}
     \caption{\small{System Model.}}
\label{fig:system model}
\vspace*{-3mm}
\end{figure}

As shown in Fig. \ref{fig:system model}, one multi-antenna BS, i.e., BS $0$, serves one single-antenna user, i.e., user $0$, with the help of one multi-element IRS in its cell, in the presence of $K$ interference BSs, i.e., BS $1$, $...$ BS $K$. Denote $\mathcal{K} \triangleq \{0,1,...,K\}$. For all $k \in \mathcal{K}\backslash\{0\}$, BS $k$ has one user, i.e., user $k$. Suppose that for all $k \in \mathcal{K}\backslash\{0\}$, the IRS is far from either BS $k$ or user $k$. Thus, each BS $k \in \mathcal{K}\backslash\{0\}$ serves user $k$, ignoring the effect of the IRS. We do not consider cooperation or coordination among the $K+1$ BSs. \blue{We consider a time period consisting of $S$ slots (coherence blocks) during which the locations of the BSs and IRS are fixed, and the users are almost static.\footnote{\blue{The time period is on the minute time-scale. One slot is on the milliseconds time-scale. Thus, $S$ is roughly $\frac{60}{0.001}=6\times10^4$.}}} Each BS $k\in\mathcal{K}$ is equipped with a uniform rectangular array (URA) of $M_k\times N_k$ antennas, and the IRS is equipped with a URA of $M_r\times N_r$ reflecting elements. For notation simplicity, define $\mathcal M_k \triangleq \{1,2,...,M_k\}$, $\mathcal N_k \triangleq \{1,2,...,N_k\}$, $\mathcal M_r \triangleq \{1,2,...,M_r\},$ and $\mathcal N_r \triangleq \{1,2,...,N_r\}$, where $k \in \mathcal{K}$. The phase shifts of the IRS's reflecting elements can be determined by a smart controller attached to the IRS. BS $0$ communicates to the IRS controller to configure the IRS's phase shifts so that the IRS can assist its communication to user $0$. In this paper, we do not consider BS cooperation or coordination.

\setcounter{TempEqCnt}{\value{equation}}
\setcounter{equation}{6}
\begin{figure*}[t]
\begin{align}
C(\mathbf{v},\mathbf{w}_0)=\mathbb{E}_{\hat{\mathbf{H}}_0,\Delta\mathbf{H}_0}\left[\log_2\left(1+
\frac{{P_0}\left\lvert\left(\mathbf{v}^H\mathbf{G}_{0,0}+ \mathbf{h}^H_{0,0}\right)\mathbf{w}_{0}\left(\hat{\mathbf{H}}_0\right)  \right\rvert^2}{\sum\limits_{k \in \mathcal{K}\backslash\{0\}} {P_{k}}\mathbb{E} \left[ {\left\lvert \left(\mathbf{v}^H\mathbf{G}_{k,0} + \mathbf{h}^H_{k,0}\right)  \frac{{\mathbf{h}}_{k,k}}{\left\lvert \left\lvert {\mathbf{h}}_{k,k} \right\rvert\right \rvert_2}\right\rvert}^2 \right]+ {\sigma}^2}\right)\right]\label{eq:sinr}
\end{align}
\hrulefill
\vspace{-1mm}
\end{figure*}
\setcounter{equation}{\value{TempEqCnt}}
We consider a narrow-band system and adopt the block-fading model for small-scale fading. As scattering is often rich near the ground, we adopt the Rayleigh fading model for the small-scale fading channels between the BSs and the users. Let $\mathbf{h}^H_{k,j} \in \mathbb{C}^{ 1 \times M_kN_k}$ denote the random channel vector for the channel between BS $k\in\mathcal{K}$ and user $j\in\mathcal{K}$ in each slot. Specifically,
\begin{align*}
\mathbf{h}^H_{k,j}=\sqrt{{\alpha_{k,j}}}\tilde{\mathbf{h}}^H_{k,j},\
\ k \in \mathcal{K}, j \in \mathcal{K},
\end{align*}
where $\alpha_{k,j}>0$ represents the large-scale fading power, and the elements of $\tilde{\mathbf{h}}^H_{k,j}$ are independent and identically distributed (i.i.d.) according to $\mathcal{CN}(0,1)$. 
As scattering is much weaker far from the ground, we adopt the Rician fading model for the small-scale fading channels between the BSs and the IRS and the small-scale fading channel between the IRS and user $0$ \cite{SJin2,YuhangJia}. Let $\mathbf{H}_{k,r} \in \mathbb{C}^{M_rN_r \times M_kN_k}$ and $\mathbf{h}^{H}_{r,0} \in \mathbb{C}^{1 \times M_rN_r}$ denote the channel matrices for the channel between BS $k\in\mathcal{K}$ and the IRS and the channel between the IRS and user $0$, respectively, in each slot. Specifically, we have:
\begin{align*}
\mathbf{H}_{k,r}=&\sqrt{\alpha_{k,r}}\! \left(\!\sqrt{\frac{K_{k,r}}{K_{k,r}+1}} \bar{\mathbf{H}}_{k,r}\! +\! \sqrt{\frac{1}{K_{k,r}+1}} \tilde{\mathbf{H}}_{k,r}\!\right)\!, k\in\mathcal{K},\\
\mathbf{h}^{H}_{r,0} = & \sqrt{\alpha_{r,0}} \left(\sqrt{\frac{K_{r,0}}{K_{r,0}+1}} \bar{\mathbf{h}}^{H}_{r,0} + \sqrt{\frac{1}{K_{r,0}+1}} \tilde{\mathbf{h}}^{H}_{r,0}\right),
\end{align*}
where $\alpha_{k,r}$, $\alpha_{r,0}>0$ represent the large-scale fading powers, $K_{k,r}$, $K_{r,0}\geq 0$ denote the Rician factors,\footnote{If $K_{k,r}=0$ or $K_{r,0}=0$, the corresponding Rician fading reduces down to Rayleigh fading. If $K_{k,r} \rightarrow \infty$ or $K_{r,0} \rightarrow \infty$, only the LoS components exist.} \blue{$\bar{\mathbf{H}}_{k,r} \in \mathbb{C}^{M_rN_r \times M_kN_k}$, $\bar{\mathbf{h}}^{H}_{r,0}\in \mathbb{C}^{1\times M_rN_r}$ represent the normalized LoS components with unit-modulus elements, and $\tilde{\mathbf{H}}_{k,r} \in \mathbb{C}^{M_rN_r \times M_kN_k}$, $\tilde{\mathbf{h}}^{H}_{r,0} \in \mathbb{C}^{1 \times M_rN_r}$ represent the random normalized NLoS components in each slot with elements i.i.d. according to $\mathcal{CN}(0,1)$.} \blue{We suppose that the channel model, i.e., all large-scale fading powers, Rician factors, LoS components, and distributions of all random NLoS components for each slot, remain unchanged during the considered time period.}

\blue{We consider a quasi-static phase shift design where the phase shifts $\mathbf{v}$ do not change with the fast varying NLoS components to reduce phase adjustment cost.}
Let $\phi_{m,n} \in [0,2\pi)$ denote the phase shift of the $(m,n)$-th element of the IRS. For notation convenience, we introduce
$\mathbf{v}\triangleq\text{vec}\left(\left(e^{j\phi_{m,n}}\right)_{m \in \mathcal M_r, n \in \mathcal N_r}\right)\in \mathbb{C}^{M_rN_r\times 1}$
to represent the phase shifts. Denote $\mathcal{N}\triangleq \{1,...,M_rN_r\}$. $\mathbf{v}$ can also be expressed as $\mathbf{v}=(v_n)_{n \in \mathcal{N}}$, where $v_n\in\mathbb{C}$ satisfies:
\begin{align}
\lvert v_n \rvert= 1,\ n \in \mathcal{N}. \label{eq:phi}
\end{align}
Then, the channel of the indirect link between BS $k\in\mathcal{K}$ and user $0$ via the IRS is given by:
\begin{align}
\mathbf{h}_{r,0}^H\text{diag}\left(\mathbf{v}\right)
\mathbf{H}_{k,r}=\mathbf{v}^H\mathbf{G}_{k,0},\ k \in \mathcal{K},
\end{align}
where diag$\left(\mathbf{x}\right)$ is a diagonal matrix with the entries of $\mathbf{x}$ on its main diagonal, and $\mathbf{G}_{k,0}\triangleq \text{diag}\left(\mathbf{h}_{r,0}^H\right)\mathbf{H}_{k,r} \in \mathbb{C}^{M_rN_r\times M_kN_k}$ is referred to as the cascaded channel between BS $k\in\mathcal{K}$ and user $0$. We express the channel of each indirect link in terms of the corresponding cascaded channel, for ease of analysis. Accordingly, the LoS components of the cascaded channel are given by:
\begin{align}
\bar{\mathbf{G}}_{k,0}\triangleq \sqrt{\alpha_{k,r}\alpha_{r,0}\tau_{k}}\text{ diag}
\left(\bar{\mathbf{h}}_{r,0}^H\right)\bar{\mathbf{H}}_{k,r},\quad k \in \mathcal{K},
\end{align}
where $\tau_{k} \triangleq \frac{K_{k,r}K_{r,0}}{(K_{k,r}+1)(K_{r,0}+1)}$.
Considering both the indirect and direct links, the equivalent channel between BS $k\in\mathcal{K}$ and user $0$ in the IRS-assisted system is expressed as  $\mathbf{h}_{k,0}^H+\mathbf{G}_{k,0}^H\mathbf{v}$.

For all $k\in\mathcal{K}$, we consider linear beamforming at BS $k$ for serving user $k$. Let $\mathbf{w}_k \in \mathbb{C}^{M_kN_k \times 1}$ denote the corresponding normalized beamforming vector, where ${\left\lVert \mathbf{w}_k \right\rVert}^2_2=1$.
Thus, the signal received at user $0$ is expressed as:
\begin{align}
Y_0 \triangleq & \sqrt{P_0}\left(\mathbf{v}^H\mathbf{G}_{0,0} + \mathbf{h} ^H_{0,0}\right) \mathbf{w}_0 X_0 \nonumber \\ & + \sum_{k \in \mathcal{K}\backslash \{0\} }\sqrt{P_k}\left(\mathbf{v}^H\mathbf{G}_{k,0} + \mathbf{h}^H_{k,0}\right) \mathbf{w}_{k} X_{k} + Z,
\end{align}
where $P_k$ denotes the transmit power of BS $k\in\mathcal{K}$, $X_k\in\mathbb{C}$ is the information symbol for user $k\in\mathcal{K}$, with $\mathbb{E}\left[{\lvert X_k\rvert}^2\right] = 1$, and $Z \sim \mathcal{CN}(0,\sigma^2)$ is the additive white Gaussian noise.

In this paper, we focus on downlink transmission. We assume that BS $0$ has perfect knowledge of all large-scale fading powers, Rician factors, LoS components and distributions of all random NLoS components, \blue{as they change relatively slowly and can be estimated with high accuracy.} Furthermore, BS $0$ estimates the cascaded channel $G_{0,0}$ and direct channel $h_{0,0}$ \blue{in each slot with certain estimation errors, as instantaneous NLoS components change from slots to slots and cannot be estimated very accurately with a very limited number of pilot symbols \cite{hong2020robust}.} Let
$\hat{\mathbf{G}}_{0,0} \in \mathbb{C}^{M_rN_r\times M_0N_0}$ and $\hat{\mathbf{h}}_{0,0}\in\mathbb{C}^{M_0N_0\times 1}$ denote the estimated imperfect CSI for $\mathbf{G}_{0,0}$ and $\mathbf{h}_{0,0}$, respectively, and let $\Delta{\mathbf{G}}_{0,0} \in\mathbb{C}^{M_rN_r\times M_0N_0}$ and $\Delta{\mathbf{h}}_{0,0} \in\mathbb{C}^{M_0N_0\times 1}$ denote the corresponding estimation errors. Thus, we have:
\begin{align}
\mathbf{G}_{0,0}=\hat{\mathbf{G}}_{0,0}+\Delta{\mathbf{G}}_{0,0},\quad
\mathbf{h}_{0,0}=\hat{\mathbf{h}}_{0,0}+\Delta{\mathbf{h}}_{0,0}.
\end{align}
As in \cite{hong2020robust}, we adopt the statistical CSI error model and assume that all elements of $\Delta{\mathbf{G}}_{0,0}$ and $\Delta{\mathbf{h}}_{0,0}$ are i.i.d. according to $\mathcal{CN}(0,\delta_{1}^2)$ and $\mathcal{CN}(0,\delta_{2}^2)$, respectively. \blue{Note that it has been shown that widely used estimation methods yield zero-mean complex Gaussian distributed estimation errors.} Thus, the elements of $\hat{\mathbf{G}}_{0,0}$ and $\hat{\mathbf{h}}_{0,0}$ are independent, the $(m,n)$-th element of $\hat{\mathbf{G}}_{0,0}$, denoted by $\hat{G}_{0,0}(m,n)$, follows $\mathcal{CN}\left(\bar{G}_{0,0}(m,n),1-\delta_1^2\right)$, and the $n$-th element of $\hat{\mathbf{h}}_{0,0}$ follows $\mathcal{CN}\left(0,1-\delta_2^2\right)$. Assume that these distributions are known to BS $0$. Denote $\hat{\mathbf{H}}_0 \triangleq \left[\hat{\mathbf{h}}_{0,0},
\hat{\mathbf{G}}_{0,0}\right]\in\mathbb{C}^{M_0N_0\times (M_rN_r+1)}$ and $\Delta{\mathbf{H}}_0\triangleq \left[\Delta{\mathbf{h}}_{0,0}, \Delta{\mathbf{G}}_{0,0}\right]\in\mathbb{C}^{M_0N_0\times (M_rN_r+1)}$. Assume that in each slot user $0$ perfectly estimates the effective channel $\left(\mathbf{v}^H\mathbf{G}_{k,0}+\mathbf{h}_{k,0}^H\right)\mathbf{w}_0 \in \mathbb{C}$ and does not know  $\left(\mathbf{v}^H\mathbf{G}_{k,0}+\mathbf{h}_{k,0}^H\right)\mathbf{w}_0, k \in \mathcal{K}\backslash\{0\}$.
For all $k \in \mathcal{K}\backslash\{0\}$, assume that in each slot BS $k$ perfectly estimates $\mathbf{h}_{k,k}$.

We consider instantaneous CSI-adaptive beamforming design at BS $0$. Let
$\mathbf{w}_{0}\left(\hat{\mathbf{H}}_0\right) \in \mathbb{C}^{M_0N_0\times 1}$ denote the normalized  beamformer for BS $0$ given the  imperfectly estimated CSI $\hat{\mathbf{H}}_0$, where
\begin{align}
\left\lVert\mathbf{w}_0(\hat{\mathbf{H}}_0)\right\rVert^2_2=1.\label{eq:w}
\end{align}
We can view $\mathbf{w}_0:\mathbb{C}^{M_0N_0\times(M_rN_r+1)}\rightarrow
\mathbb{C}^{M_0N_0\times 1}$ as a vector-valued beamforming function from imperfect CSI to a beamforming vector for BS $0$.
Recall that the IRS is far from either BS $k$ or user $k$, for all $k \in \mathcal{K}\backslash\{0\}$. Thus, for all $k \in \mathcal{K}\backslash\{0\}$, to enhance the signal received at user $k$, we consider the instantaneous  CSI-adaptive maximum ratio transmission (MRT) at BS $k$ in each slot, i.e., $\frac{{\mathbf{h}}_{k,k}}{\left\lvert \left\lvert {\mathbf{h}}_{k,k} \right\rvert\right \rvert_2}$, which relies on the perfectly estimated CSI $\mathbf{h}_{k,k}$ at BS $k$.

We consider coding over a large number of slots. \blue{Then, the ergodic rate of} user $0$, $C\left(\mathbf{v},\mathbf{w}_0\right)$ (bit/s/Hz), is given by \eqref{eq:sinr}, as shown at the top of this page,\footnote{For all $k \in\mathcal{K}\backslash\{0\}$, by treating $\left(\mathbf{v}^H\mathbf{G}_{k,0} + \mathbf{h}^H_{k,0}\right) \mathbf{w}_{k}X_{k} \sim \mathcal{CN}\left(0,\mathbb{E}\left[\left\lvert\left(\mathbf{v}^H\mathbf{G}_{k,0} +\mathbf{h}^H_{k,0}\right) \mathbf{w}_{k}\right\rvert^2\right]\right)$, which corresponds to the worst-case noise, $C\left(\mathbf{v},\mathbf{w}_0\right)$ can be achieved \cite{YuhangJia}.} where $\mathbb{E}_{\mathbf{X}}\{\cdot\}$ denotes the expectation with respect to random matrix $\mathbf{X}$.
\section{Problem Formulation}
\label{sec:averagerate}
In this section, we formulate a robust optimization of instantaneous CSI-adaptive beamforming and quasi-static phase shifts for the IRS-assisted multi-cell network with imperfect CSIT and inter-cell interference. Specifically,
we aim to maximize the ergodic rate $C\left(\mathbf{v},\mathbf{w}_0\right)$ by optimizing the phase shifts $\mathbf{v}$ and beamforming function $\mathbf{w}_0$ subject to the phase shift constraints in \eqref{eq:phi} and the normalized beamforming constraints in \eqref{eq:w}.
\begin{Prob}[Robust Optimization]\label{prob:eq_delta}
\setcounter{equation}{7}
\begin{equation}\label{prob:ins_delta}
\begin{split}
\mathop{\max}_{\mathbf{v},\mathbf{w}_0}\quad &
C(\mathbf{v},\mathbf{w}_0) \\
s.t. \quad
& \eqref{eq:phi},\ \eqref{eq:w}.
\end{split}
\end{equation}
\end{Prob}
\begin{remark}[Robust Design]
An optimal solution of Problem~\ref{prob:eq_delta} adapts to the variances of the Gaussian estimation errors $\delta_1^2$ and $\delta_2^2$ and hence is robust against CSI estimation errors.
\end{remark}
\begin{remark}[Challenge for Solving Problem~\ref{prob:eq_delta}]
As the objective function does not have an analytical expression, Problem~\ref{prob:eq_delta} has to be treated as a stochastic optimization problem. $\mathbf{v}$ is constant, $\mathbf{w}_0(\hat{\mathbf{H}}_0)$ adapts to $\hat{\mathbf{H}}_0$, $C(\mathbf{v},\mathbf{w}_0)$ is non-convex in $\left(\mathbf{v},\mathbf{w}_0\right)$, and \eqref{eq:phi} is a non-convex constraint. Thus, Problem~\ref{prob:eq_delta} is actually a two-timescale stochastic non-convex optimization problem.
Moreover, the number of random variables in Problem~\ref{prob:eq_delta} is $2M_0N_0(M_rN_r+1)$, which is usually quite large. Therefore, Problem~\ref{prob:eq_delta} is very challenging.
\end{remark}
\begin{table}
\centering
\small
\begin{tabular}{|c|c|c|c|}
\hline
Problem & Problem~\ref{prob:eq_delta} & Problem~\ref{prob:approximation} & Problem~\ref{prob:theta}  \\
\hline
Random Var.  & $\hat{\mathbf{H}}_0, \Delta{\mathbf{H}}_0$ & $\hat{\mathbf{H}}_0$ & $\hat{\mathbf{H}}_0$ \\
\hline
Opt. Var. & $\left(\mathbf{v},\mathbf{w}_0\right)$ & $\left(\mathbf{v},\mathbf{w}_0\right)$ & $\mathbf{v}$
\\
\hline
Constraints  & \eqref{eq:phi}, \eqref{eq:w}& \eqref{eq:phi}, \eqref{eq:w} & \eqref{eq:phi}
\\
\hline
Timescale& Two & Two & Single \\
\hline
\end{tabular}
\caption{Comparisons of the problems.}
\label{tab:prob}
\vspace{-3mm}
\end{table}
\addtolength{\topmargin}{0.03in}
\section{Closed-form Beamforming and Approximate Phase Shift Optimization}
In this section, we obtain a closed-form beamforming design for any given phase shifts and a more tractable stochastic non-convex approximate problem only for the phase shifts. First, we simplify the objective function of Problem~\ref{prob:eq_delta}.
As in \cite{SJin2,YuhangJia}, we can obtain an upper bound of $C(\mathbf{v},\mathbf{w}_0)$ using Jensen's inequality \blue{and channel statistics}.
\begin{theorem}[Upper Bound of $C(\mathbf{v},\mathbf{w}_0)$]\label{lem:ergodic Case 1 with reflector}
\begin{align*}
 &C(\mathbf{v},\mathbf{w}_0) \leq  \log_2\left(1+
 \frac{P_0\mathbb{E}_{\hat{\mathbf{H}}_0}\left[{g}_0\left(\mathbf{v},\mathbf{w}_0
\left(\hat{\mathbf{H}}_0\right),\hat{\mathbf{H}}_0\right)\right]}
{\sum\limits_{k \in \mathcal{K}\backslash \{0\}}P_k{g}_{k}\left(\mathbf{v}\right)+\sigma^2}\right),
\end{align*}
where ${g}_0\left(\mathbf{v},\mathbf{w}_0\left(\hat{\mathbf{H}}_0\right),\hat{\mathbf{H}}_0\right)$ and ${g}_{k}\left(\mathbf{v}\right)$ are given by:
\begin{align*}
&g_0\left(\mathbf{v},\mathbf{w}_0\left(\hat{\mathbf{H}}_0\right),\hat{\mathbf{H}}_0\right)
\triangleq
\left\lvert\left(\mathbf{v}^H\hat{\mathbf{G}}_{0,0}+ \hat{\mathbf{h}}^H_{0,0}\right)\mathbf{w}_0
\left(\hat{\mathbf{H}}_0\right)\right\rvert^2 \nonumber \\ &\quad\quad\quad\quad\quad\quad\quad\quad\quad\quad\ +
\delta_{2}^2+M_rN_r\delta_1^2,\label{eq:h_sphi}\\
&g_{k}\left(\mathbf{v}\right)\triangleq
\blue{\frac{1}{M_kN_k}}\left\lvert\mathbf{v}^H
\bar{\mathbf{G}}_{k,0}
\right\rvert^2 +\alpha_{k,r}\alpha_{r,0}M_rN_r\left(1-\tau_{k}\right)
\nonumber \\ &\quad\quad\quad\ \ +\alpha_{k,0},\ k \in \mathcal{K}\backslash\{0\}.
\end{align*}
\end{theorem}

As the upper bound in Theorem~\ref{lem:ergodic Case 1 with reflector} is a good approximation of $C(\mathbf{v},\mathbf{w}_0)$, which will be seen in Fig.~\ref{fig:ergodic}, we can consider the maximization of the upper bound instead of Problem~\ref{prob:eq_delta} \cite{YuhangJia,CPan1}. $\log_2(\cdot)$ is an increasing function, so the optimization is equivalent to the following problem, which is simpler than Problem~\ref{prob:eq_delta}, as shown in Table~I.
\begin{Prob}[Approximate Problem of Problem \ref{prob:eq_delta}]\label{prob:approximation}
\begin{equation*}
\begin{split}
\mathop{\max}_{\mathbf{v},\mathbf{w}_0}\quad & \frac{P_0\mathbb{E}_{\hat{\mathbf{H}}_0}\left[{g}_0\left(\mathbf{v},\mathbf{w}_0
\left(\hat{\mathbf{H}}_0\right),\hat{\mathbf{H}}_0\right)\right]}
{\sum\limits_{k \in \mathcal{K}\backslash \{0\}}P_k{g}_{k}\left(\mathbf{v}\right)+\sigma^2} \\
s.t. \quad
& \eqref{eq:phi},\ \eqref{eq:w}.
\end{split}
\end{equation*}
\end{Prob}

Next, \blue{using Cauchy-Schwartz inequality and the structural property of  Problem~\ref{prob:approximation},} we obtain a closed-form \blue{robust} beamforming design for any given phase shifts and equivalently transform Problem~\ref{prob:approximation}, a two-timescale non-convex problem, to a single-timescale stochastic non-convex problem only for the phase shifts, as shown in Table~I.
\begin{Prob}[Equivalent Problem of Problem~\ref{prob:approximation}]\label{prob:theta}
\begin{align*}
\begin{split}
\mathop{\max}_{\mathbf{v}}\quad &
 \frac{P_0\mathbb{E}_{\hat{\mathbf{H}}_0}\left[
 {g}_0\left(\mathbf{v},\frac{\hat{\mathbf{G}}_{0,0}^H\mathbf{v}+ \hat{\mathbf{h}}_{0,0}}{\left\lVert\mathbf{v}^H\hat{\mathbf{G}}_{0,0}+ \hat{\mathbf{h}}^H_{0,0}\right\rVert_2},\hat{\mathbf{H}}_0\right)\right]}
{\sum\limits_{k \in \mathcal{K}\backslash \{0\}}P_k{g}_{k}\left(\mathbf{v}\right)+\sigma^2}
 \\
s.t. \quad
& \eqref{eq:phi}.
\end{split}
\end{align*}
\end{Prob}
\begin{theorem}[Equivalence between Problem~\ref{prob:approximation} and Problem~\ref{prob:theta}]\label{theorem:eq}
If $\mathbf{v}^{*}$ is an optimal solution of Problem~\ref{prob:theta}, then
$\left(\mathbf{v}^{*},\mathbf{w}_0^{*}\right)$ is an optimal solution of  Problem~\ref{prob:approximation}, where
\begin{align} \mathbf{w}_0^*\left(\hat{\mathbf{H}}_0\right)=\frac{\hat{\mathbf{G}}_{0,0}^H\mathbf{v}^*+ \hat{\mathbf{h}}_{0,0}}{\left\lVert\left(\mathbf{v}^{*}\right)^H\hat{\mathbf{G}}_{0,0}+ \hat{\mathbf{h}}^H_{0,0}\right\rVert_2}.\label{eq:wequivalence}
\end{align}
\end{theorem}


\blue{The} closed-form robust \blue{instantaneous CSI-adaptive} beamforming design \blue{(for given $\mathbf{v}^*$)} in \eqref{eq:wequivalence} \blue{has computational complexity $\mathcal{O}(M_0N_0M_rN_r)$ and hence} can promptly adapt to rapid CSI changes over slots.
\section{Algorithm for Phase Shift Optimization}\label{sec:algorithm}
In this section, we obtain a robust quasi-static phase shift design by solving Problem~\ref{prob:theta}. Specifically, we propose a low-complexity algorithm to obtain a KKT point of the stochastic non-convex problem using SSCA \cite{QingqingWu3,7412752}.
At each iteration $t$, we randomly generate $L$ channel samples, denoted by $\hat{\mathbf{H}}_{0,l}^{(t)}, l=1,...,L$, according to the distribution of $\hat{\mathbf{H}}_0$ given in Section~\ref{sec:system}. Based on the samples, we approximate the objective function of Problem~\ref{prob:theta} around $\mathbf{v}^{(t-1)}$ with a concave surrogate function $f^{(t)}(\mathbf{v})$, where $\mathbf{v}^{(t-1)}\triangleq\left(v^{(t-1)}_n\right)_{n\in\mathcal{N}}$ denotes the phase shifts at iteration $t-1$. The surrogate function for Problem~\ref{prob:theta} is given by:
\begin{align}
f^{(t)}(\mathbf{v})=&c_0^{(t)}
+2Re\left\{\sum\limits_{n=1}^{M_rN_r}c_{1,n}^{(t)}\left(v_n-v_n^{(t-1)}\right)\right\}
\nonumber \\ & -\tau\sum\limits_{n=1}^{M_rN_r}\left\lvert
v_n-v^{(t-1)}_n\right\rvert^2,
\label{eq:surrogate}
\end{align}
where $c_0^{(t)} \in \mathbb{C}$ and $c_{1,n}^{(t)}\in \mathbb{C}, n\in\mathcal{N}$ are updated according to:
\begin{align}
c_{0}^{(t)}=&
\rho^{(t)}\sum\limits_{l=1}^{L}
\frac{\gamma_{ub}
\left(\mathbf{v}^{(t-1)},\mathbf{w}_{0,l}^{(t)}
\left(\hat{\mathbf{H}}_0\right),
\hat{\mathbf{H}}_l^{(t)}\right)}{L}\nonumber \\ &+
\left(1-\rho^{(t)}\right)c_{0}^{(t-1)},\label{eq:average}
\end{align}
\begin{align}
c^{(t)}_{1,n}=&
\rho^{(t)}\sum\limits_{l=1}^{L}
\frac{\nabla_{v_n}\gamma_{ub}
\left(v^{(t-1)}_n,\mathbf{w}_{0,l}^{(t)}
\left(\hat{\mathbf{H}}_0\right),
\hat{\mathbf{H}}_l^{(t)}\right)}{L}\nonumber \\ & + \left(1-\rho^{(t)}\right)c^{(t)}_{1,n},\label{eq:derivative}
\end{align}
with $c_{0}^{(0)}=0$ and $c^{(0)}_{1,n}=0,n\in\mathcal{N}$. Here, $Re\{\cdot\}$ denotes the
real part of a complex number, $\tau>0$ can be any constant, the term $\tau\sum\limits_{n=1}^{M_rN_r}\left\lvert
v_n-v^{(t-1)}_n\right\rvert^2$ is used
to ensure strong concavity,
$\rho^{(t)}$ is a positive diminishing stepsize satisfying:
\begin{align*}
&\rho^{(t)}>0,\ \lim_{t\to\infty}\rho^{(t)}=0,\ \sum_{t=1}^\infty\rho^{(t)}=\infty,\ \sum_{t=1}^\infty\left(\rho^{(t)}\right)^2<\infty,
\end{align*}
and $\gamma_{ub}
\left(v^{(t-1)}_n,\mathbf{w}_{0,l}^{(t)}
\left(\hat{\mathbf{H}}_0\right),
\hat{\mathbf{H}}_l^{(t)}\right)$ is given by:
\begin{align*}
\frac{P_0
 {g}_0\left(\mathbf{v},\frac{\hat{\mathbf{G}}_{0,0}^H\mathbf{v}+ \hat{\mathbf{h}}_{0,0}}{\left\lVert
\mathbf{v}^H\hat{\mathbf{G}}_{0,0}+ \hat{\mathbf{h}}^H_{0,0}\right\rVert_2},\hat{\mathbf{H}}_0\right)}
{\sum\limits_{k \in \mathcal{K}\backslash \{0\}}P_k{g}_{k}\left(\mathbf{v}\right)+\sigma^2}.
\end{align*}
Furthermore, we converts\footnote{We shall see that converting \eqref{eq:phi} to \eqref{eq:cvxphi} does not affect the optimality, as the optimal solution obtained under \eqref{eq:cvxphi} given in Theorem \ref{theoren:phi} satisfies \eqref{eq:phi}.} the non-convex constraints in \eqref{eq:phi} to the convex constraints \cite{QingqingWu3}:
\begin{align}
\lvert v_n\rvert \leq 1,\ n\in\mathcal{N}.\label{eq:cvxphi}
\end{align}
Then, the resulting approximate convex problem at iteration $t$ is given \blue{as follows.}
\begin{Prob}[Approximation of Problem~\ref{prob:theta} at iteration $t$]\label{prob:ins_surrogate}
\begin{align*}
\bar{\mathbf{v}}^{(t)}=\mathop{\arg}\mathop{\max}_{\mathbf{v}}\quad &
f^{(t)}(\mathbf{v}) \\
s.t. \quad & \eqref{eq:cvxphi}.
\end{align*}
\end{Prob}

As Slater's condition is satisfied, strong duality holds for Problem~\ref{prob:ins_surrogate}. Thus, based on \blue{problem decomposition and} the KKT conditions, we can obtain a closed-form optimal solution of Problem~\ref{prob:ins_surrogate}.
\begin{theorem}[Optimal Solution of Problem~\ref{prob:ins_surrogate}]\label{theoren:phi}
$\bar{\mathbf{v}}^{(t)}\triangleq\left(\bar{v}_n^{(t)}\right)_{n\in\mathcal{N}}$ with $\bar{v}_n^{(t)}=\frac{\tau v_n^{(t-1)}+c_{1,n}^{(t)}}{\left\lvert\tau v_n^{(t-1)}+c_{1,n}^{(t)}\right\rvert}\in \mathbb{C},n\in\mathcal{N}$.
\end{theorem}


Then, the phase shifts $\mathbf{v}^{(t)}$ at iteration $t$ are updated according to:
\begin{align}
\mathbf{v}^{(t)}=\left(1-\omega^{(t)}\right)\mathbf{v}^{(t-1)}+
\omega^{(t)}\bar{\mathbf{v}}^{(t)},\label{eq:updatetheta}
\end{align}
where $\omega^{(t)}$ is a positive diminishing stepsize satisfying:
\begin{align*}
&\omega^{(t)}>0,\ \lim_{t\to\infty}\omega^{(t)}=0,\ \sum_{t=1}^\infty\omega^{(t)}=\infty,\\ &\sum_{t=1}^\infty\left(\omega^{(t)}\right)^2<\infty,
\ \lim_{t\to\infty}\frac{\omega^{(t)}}{\rho^{(t)}}=0.
\end{align*}
\begin{algorithm}[t]\small
    \caption{SSCA Algorithm for Problem~\ref{prob:theta}}
\begin{small}

        \begin{algorithmic}[1]
           \STATE \textbf{initialization}: Choose an arbitrary feasible point $\mathbf{v}^{(0)}$ of Problem~\ref{prob:ins_surrogate} as the initial point.\\
           \STATE \textbf{for} $t=1,...T$ \textbf{do}
           \STATE \quad Randomly generate $L$ channel samples $\hat{\mathbf{H}}_{0,l}^{(t)},l=1,...,L$ according to the distribution of $\hat{\mathbf{H}}_0$.\\
           \STATE \quad Update $c_0^{(t)}$ and $c^{(t)}_{1,n},n\in\mathcal{N}$ according to \eqref{eq:average} and \eqref{eq:derivative}, respectively.
           \STATE \quad Calculate $\bar{\mathbf{v}}^{(t)}$ according to Theorem~\ref{theoren:phi}.
           \STATE \quad Update $\mathbf{v}^{(t)}$
           according to $\eqref{eq:updatetheta}$.
           \STATE \textbf{end for}
    \end{algorithmic}\label{alg:TTS}
    \end{small}

\end{algorithm}
%

The details of the SSCA algorithm are summarized in Algorithm~\ref{alg:TTS}. Since Problem~\ref{prob:ins_surrogate} can be solved analytically, the computation complexity of Algorithm~\ref{alg:TTS} is relatively low. Specifically, the computational complexities of Step 4, Step 5, \blue{Step 6} of Algorithm~\ref{alg:TTS} are $\mathcal{O}\left(M_0N_0M_rN_r\right)$, \blue{$\mathcal{O}\left(M_0N_0M_rN_r\right)$, and $\mathcal{O}(M_rN_r)$, respectively}. Hence, Algorithm~\ref{alg:TTS} \blue{has computational complexity} $\mathcal{O}\left(\blue{T}M_0N_0M_rN_r\right)$.  \blue{The computational time for the robust quasi-static phase shifts is negligible
compared to the considered time period. Hence, the robust quasi-static phase shift designs have
low computation and phase adjustment costs.} Furthermore,
by Theorem~2 of \cite{7412752}, we know that every limit point of $\{\mathbf{v}^{(t)}\}$, generated by Algorithm~\ref{alg:TTS} \blue{when $T\rightarrow \infty$}, denoted by $\mathbf{v}^{\dag}$, is a KKT point of Problem~\ref{prob:theta}.

In sum,
by Theorem~\ref{theorem:eq} and Algorithm~\ref{alg:TTS}, we can obtain a suboptimal solution of Problem~\ref{prob:eq_delta}, i.e., $\left(\mathbf{v}^{\dag},\mathbf{w}_0^{\dag}\right)$, where $\mathbf{v}^{\dag}$ is the obtained KKT point and $\mathbf{w}_0^{\dag}=\frac{\hat{\mathbf{G}}_{0,0}^H\mathbf{v}^{\dag}+ \hat{\mathbf{h}}_{0,0}}{\left\lVert
\left(\mathbf{v}^{\dag}\right)^H\hat{\mathbf{G}}_{0,0}+ \hat{\mathbf{h}}^H_{0,0}\right\rVert_2}$.
\addtolength{\topmargin}{0.025in}
\section{Numerical Results}\label{sec:simulation}
\begin{figure}[t]
\begin{center}
\includegraphics[width=4.2cm]{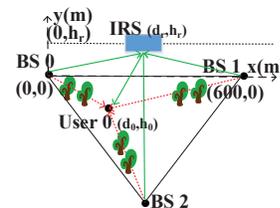}
\end{center}
\vspace{-5mm}
\caption{\small{The IRS-assisted system in Section \ref{sec:simulation}.}}
\label{fig:pathloss}
\end{figure}
\vspace{-1mm}
\begin{figure}[t]
\begin{center}
\subfigure[\scriptsize{Ergodic rate versus $M_r(=N_r)$.
}\label{fig:Ergodic_r}]
{\resizebox{4.252cm}{!}{\includegraphics{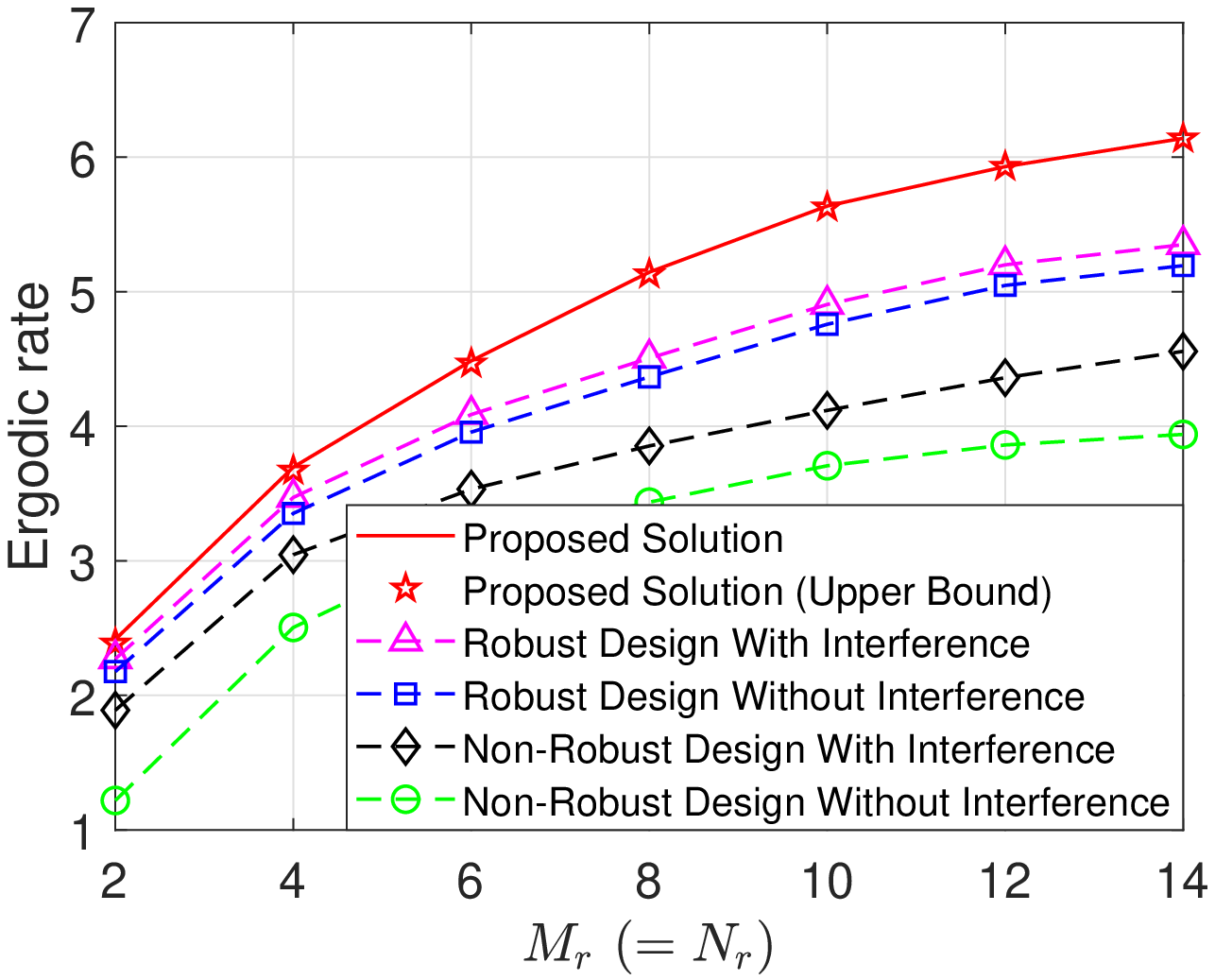}}}\quad
\subfigure[\scriptsize{Ergodic rate versus $K_{0,r}(=K_{r,0})$.
}\label{fig:Ergodic_K}]
{\resizebox{4.252cm}{!}{\includegraphics{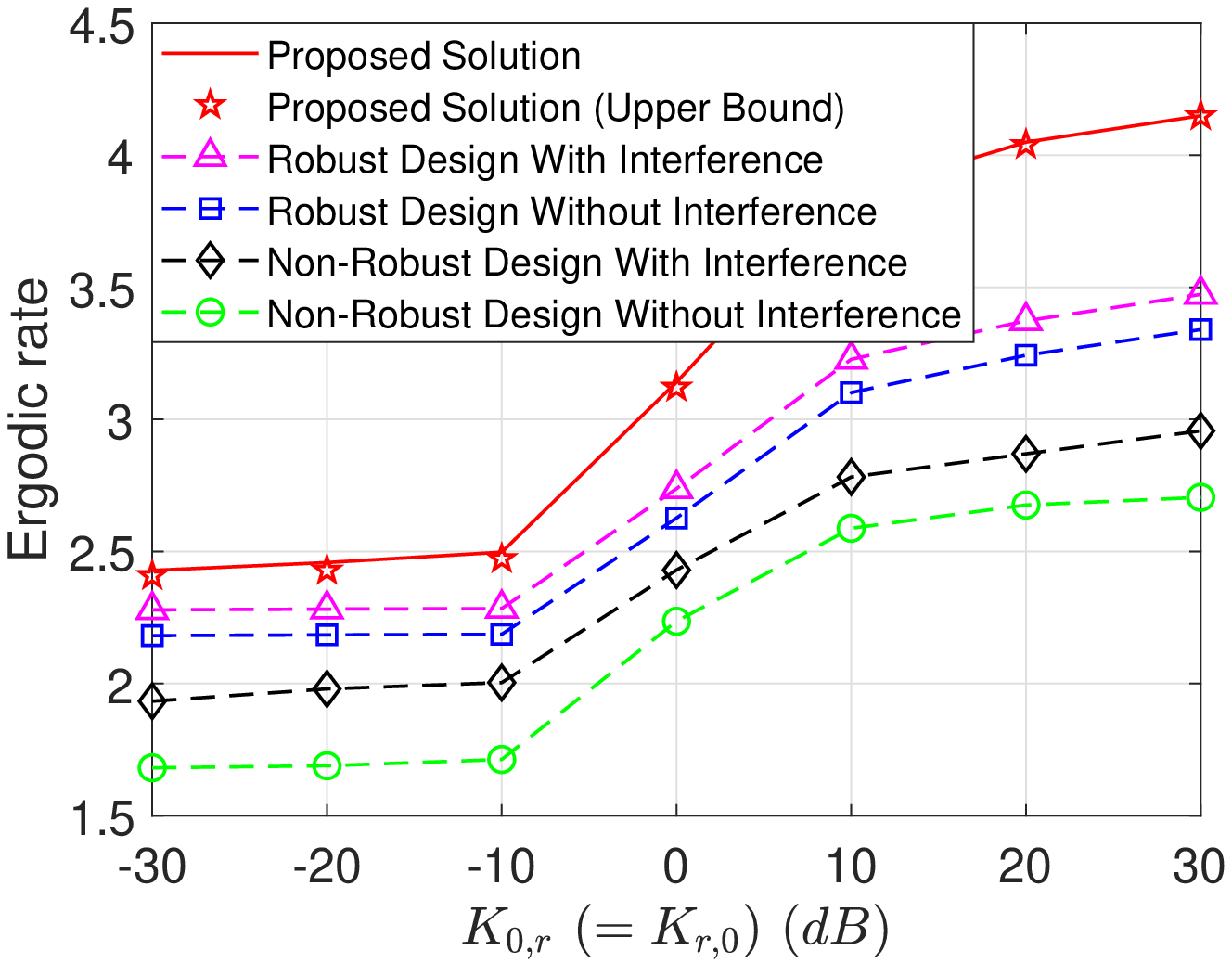}}}\quad
\subfigure[\scriptsize{Ergodic rate versus $\delta_1(=\delta_2)$.
}\label{fig:Ergodic_d_r}]
{\resizebox{4.252cm}{!}{\includegraphics{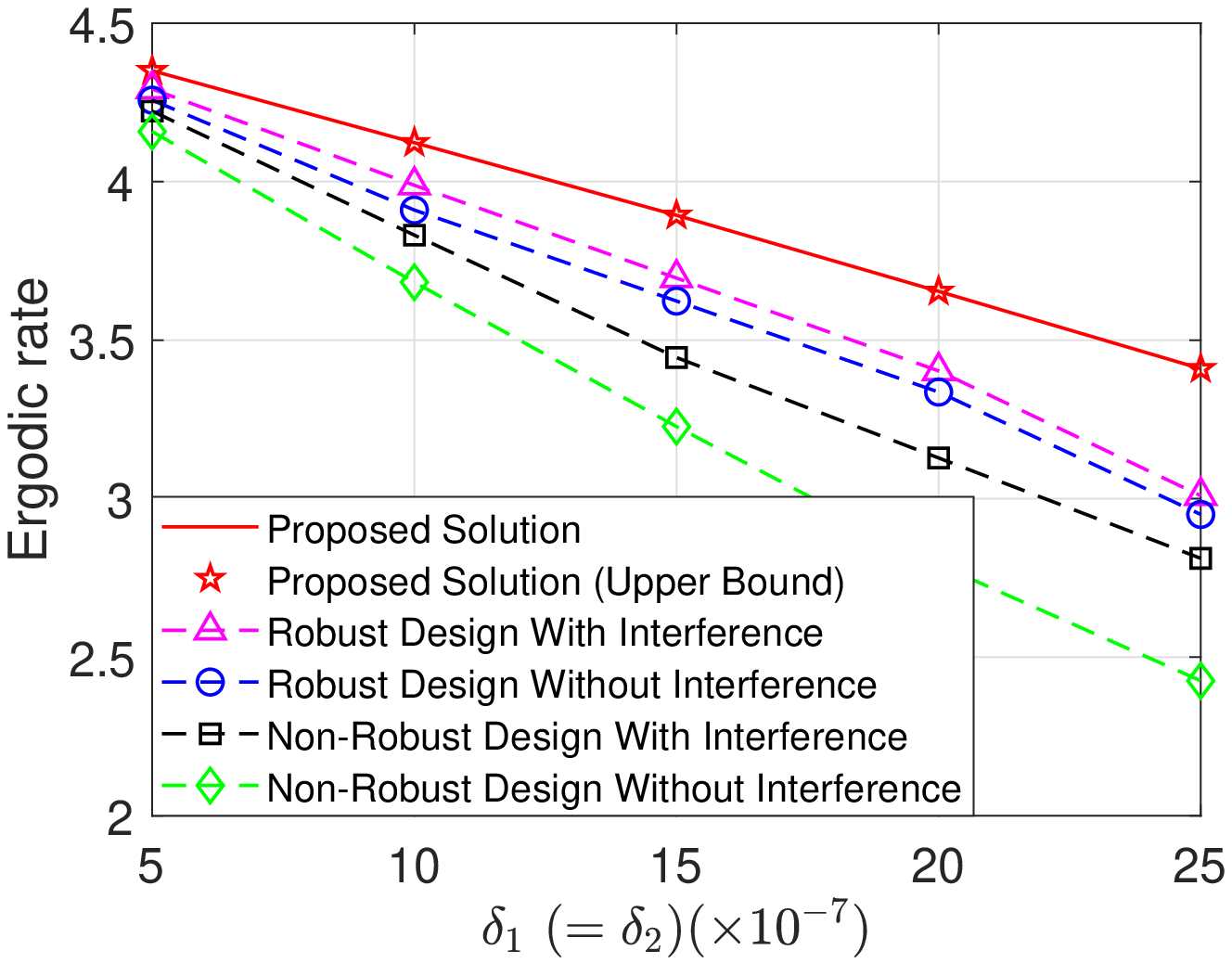}}}\quad
\subfigure[\scriptsize{Ergodic rate versus $d_{0,0}$.
}\label{fig:Ergodic_d_00}]
{\resizebox{4.252cm}{!}{\includegraphics{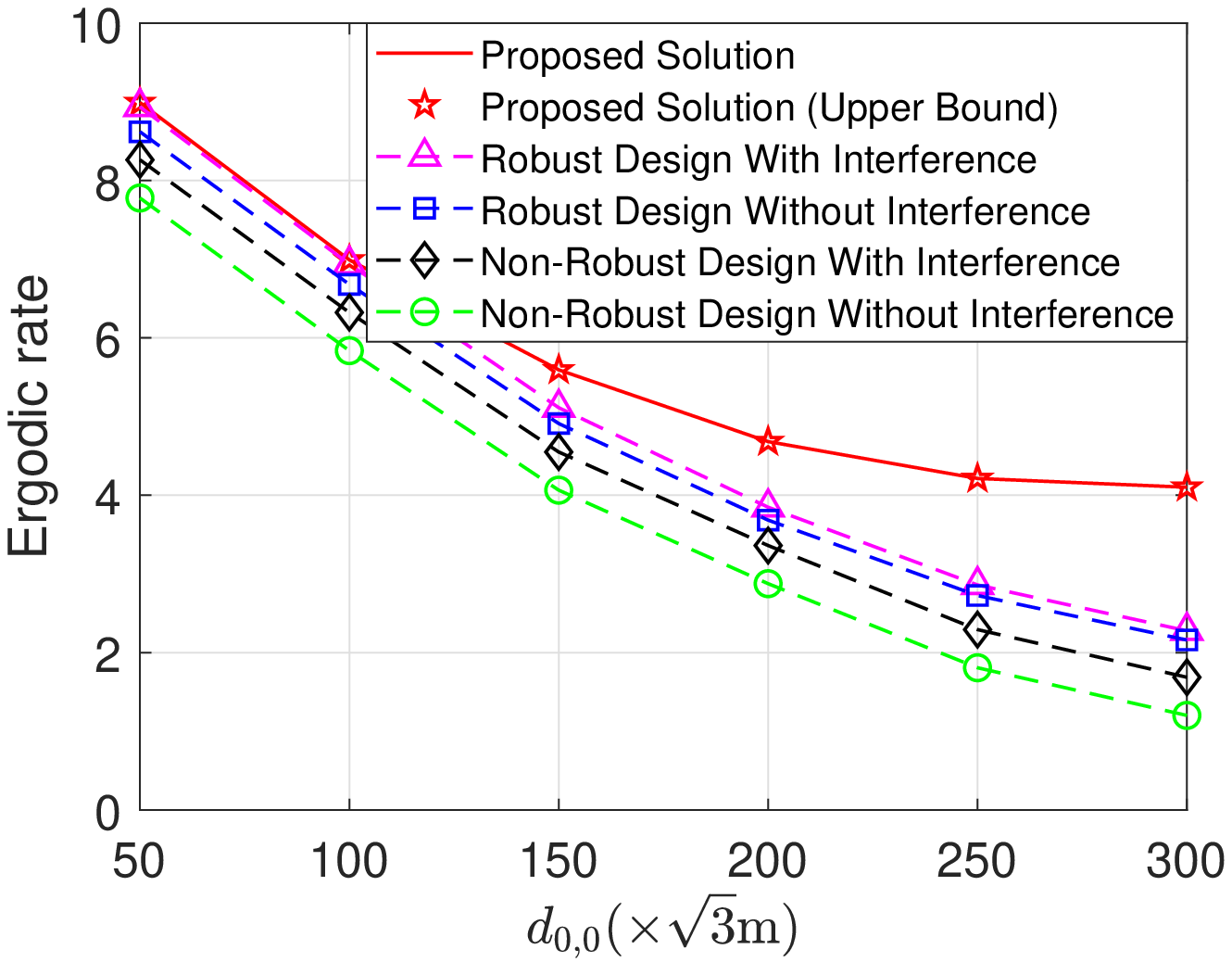}}}
\end{center}

\caption{\small{Ergodic rate.}}

\label{fig:ergodic}
\end{figure}
In this section, we numerically evaluate the proposed solutions, $\left(\mathbf{v}^{\dag},\mathbf{w}_0^{\dag}\right)$, in a multi-cell IRS-assisted system shown in Fig.~\ref{fig:pathloss}. Specifically, we  consider $K=3$. BS 0, BS 1, BS 2, user $0$, and the IRS are located at $(0,0)$, $(600,0)$, $(300,300\sqrt{3})$, $(d_{0},h_{0})$, and $(300,20)$ (in m), respectively, and the position of user $0$ lies on the perpendicular bisector of the line segment between BS $1$ and BS $2$. In the simulation, we set $d=\frac{\lambda}{2}$, $M_0=N_0=M_1=N_1=M_2=N_2=4$, $M_r=N_r=8$, $P_0=P_1=P_2=30$dBm, $\sigma^2=-90$dBm, $\varphi_{0,r}^{(h)}=\varphi_{0,r}^{(v)}=\pi/3$, $\varphi_{1,r}^{(h)}=\varphi_{1,r}^{(v)}=
\varphi_{2,r}^{(h)}=\varphi_{2,r}^{(v)}=\pi/8$, $\varphi_{r,0}^{(h)}=\varphi_{r,0}^{(v)}=\pi/6$,\footnote{$\phi_{k,r}^{(h)}
\left(\phi_{k,r}^{(v)}\right)$ represents the azimuth (elevation) angle. The definitions and notations follows those in \cite{YuhangJia}.} $d_r=250$m, $d_{0,0}=d_{1,0}=d_{2,0}=200\sqrt{3}$m, $d_{r,0}=(20+100\sqrt{3})$m, and $\delta_1=\delta_2=10^{-6}$, if not specified otherwise. We set
$\alpha_i=1/\left(1000 d_i^{\bar\alpha_i}\right)$ $(\text{i.e.,}-30+10\bar\alpha_i\log_{10}(d_i) \text{ dB}),\ i=(0,0)$, $(1,0)$, $(2,0)$, $(0,r)$, $(1,r)$, $(2,r)$, $(r,0)$, where $\bar{\alpha}_i$ represents the corresponding path loss exponent \cite{CPan1,YuhangJia}.
Due to extensive obstacles and scatters, we set $\bar\alpha_{0,0}=\bar\alpha_{1,0}=\bar\alpha_{2,0}
=3.7$. As the location of the IRS is usually carefully chosen, we assume that the links between the BSs and the IRS experience free-space path loss and set $\bar\alpha_{0,r}=\bar\alpha_{1,r}
=\bar\alpha_{2,r}=2$ \cite{YuhangJia}. In addition, we set $\bar\alpha_{r,0}=3$, due to few obstacles \cite{YuhangJia}. We consider \blue{four} baseline schemes that adopt instantaneous CSI-adaptive beamforming and quasi-static phase shift designs, namely {\em Non-Robust Design Without Interference} \cite{SJin2}, {\em Non-Robust Design With Interference} \cite{YuhangJia}, {\em Robust Design Without Interference}, \blue{and {\em Robust Design With Interference},} respectively. {\em Non-Robust Design Without Interference} and {\em Non-Robust Design With Interference} consider perfect CSIT and optimize the joint design to maximize the upper bounds on the average rates in the cases without and with inter-cell interference, respectively.

Fig. \ref{fig:ergodic} illustrates the ergodic rate versus $M_r$ $(=N_r)$, $K_{0,r}$ $(=K_{r,0})$, $\delta_1(=\delta_2)$, and $d_{0,0}$, respectively. The ergodic rates of all schemes are obtained by averaging over 10000 realizations of the random NLoS components. Fig.~\ref{fig:Ergodic_r} shows that the ergodic rate of each scheme increase with $M_r$ $(=N_r)$, mainly due to the increment of reflecting signal power. Fig.~\ref{fig:Ergodic_K} shows that the ergodic rate of each scheme increase with $K_{0,r}$ $(=K_{r,0})$, mainly due to the increment of the channel power of each LoS component. Fig.~\ref{fig:Ergodic_d_r} shows that the ergodic rate of each scheme decrease with $\delta_1(=\delta_2)$, due to the increment of channel estimation error, and the ergodic rates of the two non-robust designs decrease faster than those of the two robust ones. Fig.~\ref{fig:Ergodic_d_00} shows that the ergodic rate of each scheme decrease with $d_{0,0}$, mainly due to the increment of interference.

The Monte Carlo result and the analytical result of the proposed solution are very close to each other, indicating that the upper bound in Theorem~\ref{lem:ergodic Case 1 with reflector} is a good approximation of $C(\mathbf{v},\mathbf{w}_0)$. \blue{The gain of the proposed solution
over {\rm Robust Design With Interference} stems from the joint design of instantaneous CSI-adaptive beamforming and quasi-static phase shift design;}
the gain of the proposed solution over {\em Robust Design Without Interference} comes from the effective utilization of inter-cell interference; the gain of the proposed solution over {\em Non-Robust Design With Interference} comes from the effective utilization of imperfect CSIT; the gain of the proposed solution over {\em Non-Robust Design Without Interference} comes from the effective utilization of imperfect CSIT and inter-cell interference. The \blue{three} robust designs outperform the two non-robust designs, demonstrating the advantage of explicitly considering CSI \blue{estimation} errors. \blue{The proposed solutions and {\rm Robust Design
With Interference} outperform {\rm Robust Design Without Interference}, and {\rm Non-Robust Design With
Interference} outperforms {\rm Non-Robust Design Without Interference}, indicating the advantage of
explicitly considering inter-cell interference.}
\section{Conclusion}
This paper investigated the robust optimization of instantaneous CSI-adaptive beamforming and quasi-static phase shifts for maximizing the ergodic rate of an IRS-assisted system in the presence of channel estimation errors and inter-cell interference. Such problem is a very challenging two-timescale stochastic non-convex problem. By statistical analysis, exploiting problem structure, and using SSCA, we obtained a practical solution which does not require computing or adjusting phase shifts in each slot and is effective under channel estimation errors and inter-cell interference. Numerical results further demonstrate notable gains of the proposed robust joint design over existing instantaneous CSI-adaptive beamforming and quasi-static phase shift designs.

\bibliographystyle{IEEEtran}
\bibliography{IEEEabrv,IEEEexample}

\vspace{12pt}

\end{document}